\documentclass[twocolumn,aps,showpacs,prb]{revtex4-1}

\usepackage{graphicx}
\usepackage{epstopdf}
\usepackage{amsmath}
\usepackage{multirow}

\begin{document}

\title{First-principles study of the crystal structure, electronic structure, and transport properties of NiTe$_2$ under pressure}

\author{Jian-Feng Zhang$^1$}
\author{Yawen Zhao$^2$}
\author{Kai Liu$^1$}\email{kliu@ruc.edu.cn}
\author{Yi Liu$^2$}\email{yiliu@bnu.edu.cn}
\author{Zhong-Yi Lu$^1$}\email{zlu@ruc.edu.cn}

\affiliation{$^1$Department of Physics and Beijing Key Laboratory of Opto-electronic Functional Materials $\&$ Micro-nano Devices, Renmin University of China, Beijing 100872, China}
\affiliation{$^2$Center for Advanced Quantum Studies and Department of Physics, Beijing Normal University, Beijing 100875, China}

\date{\today}

\begin{abstract}
Recent experiments showed the distinct observations on the transition metal ditelluride NiTe$_2$ under pressure: one reported a superconducting phase transition at 12 GPa, whereas another observed a sign reversal of Hall resistivity at 16 GPa without the appearance of superconductivity. To clarify the controversial experimental phenomena, we have carried out first-principles electronic structure calculations on the compressed NiTe$_2$ with structure searching and optimization. Our calculations show that the pressure can transform NiTe$_2$ from a layered P-3m1 phase to a cubic Pa-3 phase at $\sim$10 GPa. Meanwhile, both the P-3m1 and Pa-3 phases possess nontrivial topological properties. The calculated superconducting $T_c$'s for these two phases based on the electron-phonon coupling theory both approach 0 K. Further magnetic transport calculations reveal that the sign of Hall resistance for the Pa-3 phase is sensitive to the pressure and the charge doping, in contrast to the case of the P-3m1 phase. Our theoretical predictions on the compressed NiTe$_2$ wait for careful experimental examinations.
\end{abstract}


\maketitle

\section{INTRODUCTION}

Transition metal dichalcogenides (TMDs) have attracted extensive attention for decades since they have exhibited versatile physical phenomena, such as charge density wave (CDW), extremely large magnetoresistance, nontrivial topological band structure, magnetism, and superconductivity. The TMD materials usually crystalize in a layered structure with each layer consisting of a triangular transition metal sheet sandwiched by two chalcogen planes. In each TMD layer, the transition metal atoms have either trigonal prismatic coordination or octahedral coordination with the chalcogen atoms, i.e. the 2H and 1T phases, respectively. Because of the layered crystal structure, the physical properties of TMDs can be effectively tuned by external pressures. For example, the pressure can suppress the CDW order in 1T-Ta(Se/S)$_2$, 1T-VSe$_2$, 2H-NbSe$_2$, and 2H-Ta(Se/S)$_2$, and concomitantly induce the superconductivity\cite{1ttas2pre,1tvse2pre,2hnbse2pre,2htas2pre}. The extremely large magnetoresistance in WTe$_2$ also vanishes with increasing pressure, accompanying the appearance of superconductivity\cite{wte2pre}. Furthermore, the topological band structure, such as the type of Dirac cone in PdTe$_2$ can be modulated by the pressure\cite{pdte2pre}. The TMD materials with the rich phase diagrams under pressure, thus, provide ideal platforms to explore the related physical mechanisms.

Recently a new TMD material 1T-NiTe$_2$ has been intensively studied due to its nontrivial topological property with the type-II Dirac points around the Fermi level\cite{nite2,nite22}. Notably, there are distinct experimental observations on NiTe$_2$ under pressure\cite{presssc,presshr}. One experiment reported that NiTe$_2$ underwent a superconducting phase transition at 12 GPa~\cite{presssc}. In contrast, another experiment showed that NiTe$_2$ had a sign reversal of Hall resistivity at 16 GPa but without the emergence of superconductivity \cite{presshr}. These interesting phenomena have triggered our curiosity: What is the mechanism underlying the superconductivity or the sign reversal of Hall resistivity observed in the compressed NiTe$_2$?

To study the effect of pressure on NiTe$_2$, we first performed the structure searching calculations on NiTe$_2$ under high pressures. We found that above 10 GPa a new cubic phase with space group Pa-3 has lower enthalpy than the well-known layered phase with space group P-3m1, indicating the possibility of pressure-induced structural phase transition that has not been reported before~\cite{presssc,presshr}. We then studied the topological property, superconductivity, and Hall resistivity for both the P-3m1 and the Pa-3 phases of NiTe$_2$ under pressure, and analyzed the relations between our calculations and previous experiments~\cite{presssc,presshr}.

\section{Method}

To investigate the crystal structures and the electronic structures of NiTe$_2$ under high pressure, we carried out the density functional theory\cite{dft1,dft2} (DFT) calculations by using the Vienna Ab-initio simulation package (VASP)\cite{vasp1,vasp2} in combination with the swarm-intelligence-based CALYPSO structure prediction method\cite{calypso1,calypso2}. Six independent searching missions under 50 GPa (respective cell sizes of 1 - 4, 6, and 8 formula units) were performed, and more than 600 structures in each mission were produced. The generalized gradient approximation of Perdew-Burke-Ernzerhof \cite{PBE} type was adopted for the exchange-correlation functional. The optB86b functional was employed to describe the interlayer van der Waals (vdW) interactions \cite{optb86b}. A kinetic-energy cutoff of 350 eV was used for the plane-wave basis. The Gaussian smearing method with a width of 0.05 eV was used for the Fermi-surface broadening. A Monkhorst-Pack $\bf {k}$-point mesh with a grid spacing of 0.2 \AA$^{-1}$ was adopted for the Brillouin-zone (BZ) sampling. %To examine the convergence of enthalpy calculations,
We also studied several structures with the lowest enthalpies in each mission by using a dense $\bf {k}$-point grid with a spacing of 0.05 \AA$^{-1}$ and higher kinetic-energy cutoffs (450 and 550 eV) to ensure that the error bars of our enthalpy calculations are below the 1 meV/formula. Comparing with the mainly studied P-3m1 and Pa-3 phases, all of the other searched structure phases own higher enthalpies in the pressure range of 0 - 50 GPa.
%The surface states in the two-dimensional (2D) BZ were studied by using WannierTools package \cite{wt}.
In phonon calculations, the real-space force constants were calculated within the density functional perturbation theory \cite{dfpt} implemented in the VASP package and the phonon dispersion was then obtained with the PHONOPY code\cite{phonopy}. To simulate the x-ray diffraction (XRD) spectrum, the x-ray wavelength was set to 0.4133 \AA, the same as that in Ref. \onlinecite{presshr}.
In the study of the topological properties for the P-3m1 phase NiTe$_2$, the parities of the wave functions at the time-reversal invariant momentum (TRIM) points can be obtained by a point-group analysis of these wave functions as implemented in the Quantum ESPRESSO (QE)\cite{pwscf} package. As for the Pa-3 phase with a nonsymmorphic space group, we distinguished the parities of the wave functions by analyzing the corresponding orbital compositions of the Ni atom with the center inversion.

The superconductivity of NiTe$_2$ under high pressure was studied based on the electron-phonon coupling (EPC) theory with the ELECTRON-PHONON WANNIER software\cite{epw}, which is interfaced with the QE \cite{pwscf} and WANNIER90 packages\cite{mlwf}. For the P-3m1 (Pa-3) phase, we adopted the 6$\times$6$\times$4 (6$\times$6$\times$6) {\bf k}-mesh and the 6$\times$6$\times$4 (2$\times$2$\times$2) {\bf q}-mesh as coarse grids and interpolated to the 48$\times$48$\times$32 (36$\times$36$\times$36) {\bf k}-mesh and the 24$\times$24$\times$16 (18$\times$18$\times$18) {\bf q}-mesh dense grids. The superconducting transition temperature $T_c$ was calculated by substituting the EPC constant $\lambda$ into the McMillan-Allen-Dynes formula \cite{mcmillan1, mcmillan2}, ${T_c=\frac{\omega_{log}}{1.2}\text{exp}[\frac{-1.04(1+\lambda)}{\lambda(1-0.62\mu^*)-\mu^*}]}$, where $\omega_{log}$ is the logarithmic average of the Eliashberg spectral function\cite{Eliashberg} and $\mu^*$ is the effective screened Coulomb repulsion constant that was set to an empirical value of 0.1\cite{mustar1,mustar2}.

The magnetic transport properties of NiTe$_2$, especially the conductivity tensor $\hat{\sigma}$ in magnetic-field ($\bf {B}$), was studied based on the Boltzmann transport equation with the relaxation-time approximation\cite{bte}: ${\hat{\sigma}}^n({\bf B})=\frac{e^2}{4\pi^3}\int{d{\bf k}\tau_{n}({\bf k}){\bf v}_{n}({\bf k}){\bar{\bf v}}_{n}({\bf k})(-\frac{\partial f}{\partial \varepsilon})_{\varepsilon=\varepsilon_{n}({\bf k})}}$, where $n$ is the band index, ${\bf v}_{n}({\bf k})$ is the velocity given by the equation of motion: ${\bf v}_{n}({\bf k})=\frac{1}{\hbar}\frac{\partial \varepsilon_{n}({\bf k})}{\partial {\bf k}}$, ${\bar {\bf v}}_{n}({\bf k})$ is the weighted average of the velocity over the past history of the electron passing through ${\bf k}$, and ${\bar {\bf v}}_{n}({\bf k})=\int_{-\infty}^{0}\frac{dt}{\tau_{n}({\bf k})}e^{t/\tau_{n}({\bf k})}{\bf v}_{n}[{\bf k}_n(t)]$. The magnetic field is explicitly acting on the time evolution of ${\bf k}_n(t)$, as given in the equation of motion: $\frac{d{\bf k}_n(t)}{dt}=-\frac{e}{\hbar}{\bf v}_{n}[{\bf k}_n(t)]\times{\bf B}$ with the initial condition ${\bf k}_n(0)={\bf k}$, $f$ is the Fermi-Dirac distribution, and ${\varepsilon}$ is the electron energy. The calculations were carried out with the WANNIERTOOLS package\cite{wt} that is interfaced to the WANNIER90 package\cite{mlwf}. In the transport calculations, the electronic relax-time $\tau_{n}({\bf k})$ was assumed as an unknown constant $\tau$, and the relation between $\hat{\sigma}/\tau$ ($\hat{\rho}\tau$) and ${\bf B}\tau$ was computed\cite{liuyi}. A 101$\times$101$\times$101 $\bf {k}$-mesh was employed to describe the electronic structure around the Fermi level. The directions of $\bf {B}$ were set to the [001] direction for the P-3m1 phase and to the [111] direction for the Pa-3 phase, respectively.
%For the Pa-3 phase, the influence of magnetic field direction on the transport properties had been also examined.
The Hall resistivity $R_H$ was calculated by $-\rho_{xy}\tau/{\bf B}\tau$.

\section{Results and Discussion}

\begin{figure}[tb]
\includegraphics[angle=0,scale=0.29]{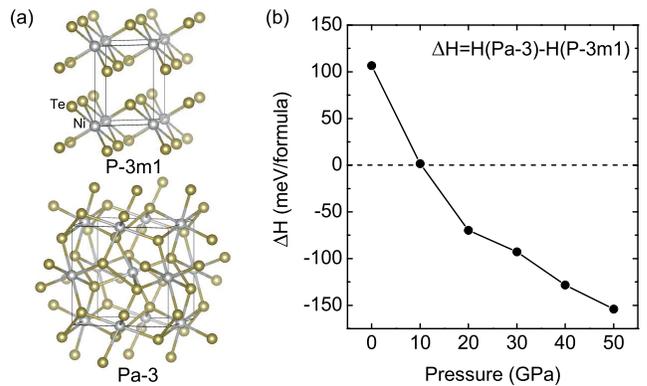}
\caption{(Color online) (a) Crystal structures of the P-3m1 and Pa-3 phases of NiTe$_2$, where the gray and yellow balls represent the Ni and Te atoms, respectively. (b) Pressure dependent enthalpy differences ($\Delta H$) between the P-3m1 and Pa-3 phases of NiTe$_2$: $\Delta H$ = $H(\text{Pa-3}) - H(\text{P-3m1})$.}
\label{fig1}
\end{figure}

$Structural\ transition:$ Previous experiments show that NiTe$_2$ adopts the P-3m1 phase at ambient pressure [upper panel in Fig. 1(a)]\cite{nite2,nite22}. It is a two-dimensional structure with the Ni atomic layer sandwiched by two layers of Te atoms. Applying high pressure, we found a stable pyrite structure, named the Pa-3 phase [lower panel in Fig. 1(a)], based on structure searching calculations. The Ni atoms therein form a face-centered-cubic lattice, and each Ni atom is surrounded by six Te atoms. The spacial inversion symmetry holds in both the P-3m1 and Pa-3 phases. It is worth mentioning that the Pa-3 phase is found in experiment for NiS$_2$ and NiSe$_2$ at ambient pressure\cite{nise2}. Figure \ref{fig1}(b) shows the pressure dependence of the enthalpy differences between the Pa-3 and the P-3m1 phases $\Delta H$ = $H(\text{Pa-3}) - H(\text{P-3m1})$. The P-3m1 phase is energetically more stable than the Pa-3 phase at low pressure, in consistence with the previous observations\cite{nite2,presssc,presshr}. Once the applied pressure exceeds 10 GPa, $\Delta H$ experiences a sign reversal, indicating a structural transition from the P-3m1 phase to the Pa-3 phase. This has not been reported in the previous experiments\cite{presssc,presshr}. Considering the vdW correction may not apply to the three-dimensional structure of the Pa-3 phase, we also checked our calculations with the DFT-D3 correction~\cite{dftd3} and without the vdW correction, and still get the sign reversal of $\Delta H$ around 10 GPa.

\begin{figure}[tb]
\includegraphics[angle=0,scale=0.4]{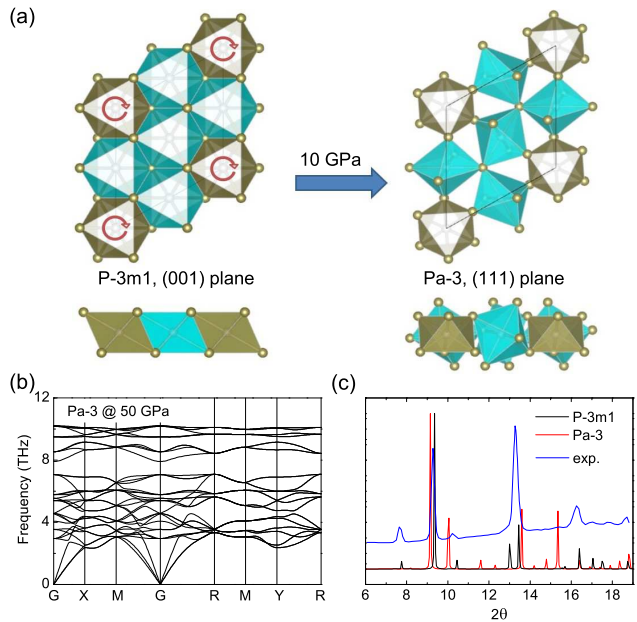}
\caption{(Color online) (a) Schematic structural transition process from the P-3m1 phase to the Pa-3 phase. The displayed structures are the top views (upper panels) and side views (bottom panels) of the (001) plane of the P-3m1 phase and the (111) plane of the Pa-3 phase. Monolayers of these two phases are displayed for clarity. (b) Phonon dispersion for the Pa-3 phase of NiTe$_2$ at 50 GPa. (c) Predicted XRD spectra of NiTe$_2$ in the P-3m1 phase (black line) and the Pa-3 phase (red line) at 50 GPa. The experimental XRD spectrum at 52.2 GPa (blue line) (Ref. \onlinecite{presshr}) is also exhibited for comparison.}
\label{fig2}
\end{figure}

Figure \ref{fig2}(a) schematically shows a possible structural transition process between the P-3m1 and Pa-3 phases. The displayed images are the monolayers for the (001) plane of the P-3m1 phase and the (111) plane of the Pa-3 phase, respectively.
%It should be noted that the Pa-3 phase of NiTe$_2$ is not of vdW type.
The former can be transformed into the latter when the Ni-Te octahedra around the corresponding vertices [yellow octahedra in Fig. 2(a)] rotate by about 30$^{\circ}$ (red arrows), and some related Ni-Te bonds break up. The Ni atoms with the broken bonds then bond with the Te atoms in other atomic layers and form new Ni-Te octahedra [blue octahedra in Fig. 2(a)].
%This implies that the transition barrier between these two phases may be low.
%In fact, the Pa-3 phase is also the experimentally determined structure of the nickel chalcogenides NiS$_2$ and NiSe$_2$ at ambient pressure \cite{nise2}.
Nevertheless, the previous measurements did not report any structural phase transition in the XRD spectrum of NiTe$_2$ under pressure~\cite{presssc,presshr}, which, thus, calls for further experimental examination.

To inspect the structural stability of the predicted Pa-3 phase of NiTe$_2$ under pressure, we calculated its phonon dispersion at a typical pressure of 50 GPa [Fig. \ref{fig2}(b)]. There is no imaginary frequency in the whole BZ, indicating its dynamical stability. We also simulated the XRD spectrum for the Pa-3 phase of NiTe$_2$ at the same pressure [Fig. \ref{fig2}(c)]. For comparison, the simulated XRD spectrum of the P-3m1 phase is also presented. As can be seen, most peaks of the Pa-3 phase (red line) are very close to those of the P-3m1 phase (black line). In fact, by careful examination, the peaks of the Pa-3 phase can be also found in the experimental data (blue line) of Ref. \onlinecite{presshr}. This means that the realistic NiTe$_2$ compound under high pressure may be a mixture of the P-3m1 and Pa-3 phases, i.e. these two phases distributing separately in different areas.
The calculated lattice parameters for the Pa-3 phase of NiTe$_2$ are listed in Table I, which can serve as a reference for future experiment.

\begin{figure}[tb]
\includegraphics[angle=0,scale=0.39]{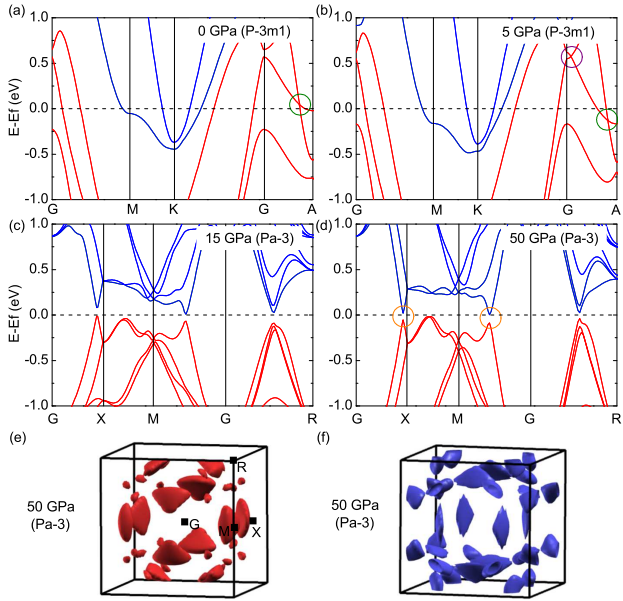}
\caption{(Color online) Band structures of NiTe$_2$ at (a) 0 GPa, (b) 5 GPa, (c) 15 GPa, and (d) 50 GPa in corresponding stable phases. The spin-orbit coupling (SOC) is included. The red and blue lines represent the valence and conduction bands with respect to a curved Fermi level, respectively.
%For the P-3m1 phase, the type-II Dirac point at ambient pressure and the type-I Dirac points at 5 GPa are highlighted by blue circles. For the Pa-3 phase, the Dirac points gapped by the SOC effect are highlighted by blue circles.
(e) Hole-type and (f) electron-type Fermi surfaces of NiTe$_2$ in the Pa-3 phase at 50 GPa.}
\label{fig3}
\end{figure}

\begin{table}[!b]
\caption{Lattice parameters and atomic Wyckoff positions for the Pa-3 phase of NiTe$_2$ under 15 and 50 GPa.}
\begin{center}
\begin{tabular*}{8cm}{@{\extracolsep{\fill}} cccc}

\hline \hline
Pressure (GPa) & $a$ (\AA) & Ni, 4$b$ & Te, 8$c$ \\
\hline
15 & 6.11 &  (0.50, 0.00, 0.00) & (0.64, 0.14, 0.36) \\
50 & 5.79 &  (0.50, 0.00, 0.00) & (0.64, 0.14, 0.36) \\
\hline\hline

\end{tabular*}
\end{center}
\end{table}

$Band\ topology:$ We then studied the band structures and topological properties of NiTe$_2$ in the corresponding stable phases under different pressures. The band structures of NiTe2 calculated with the SOC effect at 0 GPa (P-3m1), 5 GPa (P-3m1), 15 GPa (Pa-3), and 50 GPa (Pa-3) are shown in Figs. 3(a)-3(d), respectively. The red and blue lines represent the valence and conduction bands separated by a curved Fermi level. Since NiTe$_2$ has both time-reversal and space-inversion symmetries in both the P-3m1 and the Pa-3 phases, we can calculate the corresponding topological invariants Z$_2$ by the product of the parities of all occupied bands (red bands in Fig. 3) at the eight TRIM points \cite{fu2007prb}. As shown in Table II, the Z$_2$ invariants of NiTe$_2$ maintain one with applied pressure, revealing its nontrivial topological properties before and after the structural phase transition. The topological properties of both phases were also checked by the Wilson loop method as implemented in the WANNIERTOOLS package~\cite{wt}, and the same conclusion was obtained.

\begin{table}[!b]
\caption{Total parities of all occupied bands below the full gap (red bands in Fig. 3) at eight TRIM points in the BZ of NiTe$_2$.}
\begin{center}
\begin{tabular*}{8cm}{@{\extracolsep{\fill}} cccccc}
\hline \hline
P-3m1 (GPa) & G & 3M & A & 3L & Total\\
\hline
0 & + & - & + & + & -\\
5 & + & - & + & + & -\\
\hline \hline
Pa-3 (GPa) & G & 3X & 3M & R & Total\\
\hline
15 & - & + & + & + & -\\
50 & - & + & + & + & -\\
\hline\hline
\end{tabular*}
\end{center}
\end{table}

The nontrivial topological properties of NiTe$_2$ can be learned in more detail from its band structure. In the P-3m1 phase at ambient pressure, NiTe$_2$ has a type-II Dirac point along the G-A path of the BZ slightly above the Fermi level\cite{nite2,nite22}, which is highlighted by a green circle in Fig. 3(a). At low pressure, for example 5 GPa, this type-II Dirac point moves below the Fermi level [Fig. 3(b)], implying that the pressure can effectively tune the position of the type-II Dirac point. Moreover, a type-I Dirac point can be induced by the pressure along the G-A path [labeled by a purple circle in Fig. 3(b)].
%Similar variations were also predicted in other congener element compounds (PdTe$_2$ and PtTe$_2$) (who predicts??).
Note that this type-I Dirac point is generated by the band crossing of two valence bands, which does not influence the Z$_2$ invariant of NiTe$_2$, as shown in Table II.

The band structures for the Pa-3 phase of NiTe$_2$ at 15 and 50 GPa are shown in Figs. 3(c) and 3(d), respectively. Although no band crosses the Fermi level along the high-symmetry paths in the BZ, the calculated Fermi surfaces at 50 GPa [Figs. 3(e) and 3(f)] indicate that there are multiple small Fermi pockets in the interior of the BZ. Along the G-X and G-M paths of the BZ, there are Dirac-type band crossings that are gapped by the SOC effect [as highlighted by the orange circles in Fig. 3(d)]. In fact, there is a full energy gap between the valence bands (red color) and the conduction bands (blue color) for the Pa-3 phase in the whole BZ. In consideration of the nontrivial Z$_2$ invariant (Table II), the Pa-3 phase of NiTe$_2$ can be also regarded as a topological insulator defined on a curved Fermi level.

$Superconductivity:$
%One important experimental finding is NiTe$_2$ will undergo a superconducting transition at 12 GPa with 6 K's superconducting Tc \cite{presssc}. In addition, the monolayer NiTe$_2$ was also predicted to be a superconductor with 5.7 K's superconducting $T_c$ \cite{mononite2}.
%But in another experiment of pressed NiTe$_2$, the superconductivity always absent with pressure \cite{presshr}.
To examine the possible superconductivity of NiTe$_2$ under pressure reported by the previous experiment \cite{presssc}, we performed the EPC calculations on the P-3m1 and the Pa-3 phases of NiTe$_2$ at 50 GPa.
%The SOC effect was not included.
%The superconducting $T_c$ was calculated based on the McMillan-Allen-Dynes formula\cite{mcmillan1, mcmillan2}.
The calculated superconducting $T_c$, EPC constant $\lambda$, and logarithmic average of Eliashberg spectral function $\omega_{log}$ are listed in Table III. The $T_c$'s of the P-3m1 and Pa-3 phases of NiTe$_2$ both approach 0 K, revealing the absence of EPC-derived conventional superconductivity.
To check the influence of the empirical $\mu^*$ on $T_c$, we performed additional calculations with several values of $\mu^*$ ranging from 0.00 to 0.14. The calculated superconducting $T_c$ is always below 0.4 K, which suggests that the EPC even with the addition of the plasmonic effect (equivalent to reducing $\mu^*$ and enhancing the mass)~\cite{plasmon} cannot provide the reported superconductivity ($T_c\sim$ 3.7-7.5 K) on polycrystalline NiTe$_2$~\cite{presssc}.
In addition, we also performed a spin-polarized calculation on NiTe$_2$ and obtained a nonmagnetic ground state, which suggests the absence of spin-fluctuation-induced unconventional superconductivity either. We thus deduce that the previously observed superconductivity in polycrystalline NiTe$_2$~\cite{presssc} may not be an intrinsic property. In comparison, the measurements on NiTe$_2$ single crystals revealed the absence of superconductivity under pressure~\cite{presshr}, which are consistent with our calculations.

\begin{table}[!b]
\caption{The calculated superconducting transition temperature $T_c$, EPC constant $\lambda$, and logarithmic average of Eliashberg spectral function $\omega_{log}$ of NiTe$_2$ in the P-3m1 and Pa-3 phases at 50 GPa.}
\begin{center}
\begin{tabular*}{8cm}{@{\extracolsep{\fill}} cccc}
\hline \hline
  & $T_c$ (K) & $\lambda$ & $\omega_{log}$ (cm$^{-1}$)\\
\hline
P-3m1 & 0.02 & 0.26 & 112.9 \\
Pa-3 & 0.00 & 0.21 & 127.6 \\
\hline\hline
\end{tabular*}
\end{center}
\end{table}

\begin{figure}[tb]
\includegraphics[angle=0,scale=0.28]{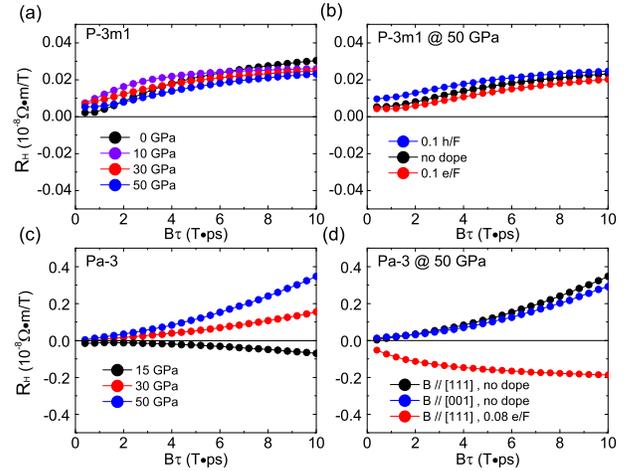}
\caption{(Color online) Calculated Hall resistance $R_\text{H}$ of (a) the P-3m1 phase and (c) the Pa-3 phase of NiTe$_2$ under different pressures. (b) Influence of charge doping on $R_\text{H}$ for the P-3m1 phase of NiTe$_2$ at 50 GPa. The h/F (e/F) represent holes (electrons) per formula unit. (d) Influence of charge doping and magnetic field direction on $R_\text{H}$ for the Pa-3 phase of NiTe$_2$ at 50 GPa.}
\label{fig4}
\end{figure}

$Hall\ resistance:$ Another interesting observation on NiTe$_2$ is the sign reversal of Hall resistivity under pressure\cite{presshr}, which indicates a transition of the majority carrier type\cite{presshr}. To clarify the origin of this phenomenon, we studied the Hall resistance $R_\text{H}$ of NiTe$_2$ in different crystal phases and under various pressures, doping conditions, and magnetic field directions. As the electronic transport properties are determined by the electronic band structure~\cite{liuyi}, the SOC effect was included to precisely describe the electronic states around the Fermi level.
As shown in Fig. 4(a), the sign of $R_\text{H}$ for the P-3m1 phase of NiTe$_2$ is insensitive to the applied pressure and remains positive with increasing pressure. At high pressure, for example 50 GPa, the sign of $R_\text{H}$ is also independent on the type of doping carriers [Fig. 4(b)]. These results suggest that if NiTe$_2$ remained in the P-3m1 phase and there was no structural phase transition under pressure, the Hall resistance would not change sign, which is inconsistent with the previous experiment\cite{presshr}. On the other hand, for the Pa-3 phase of NiTe$_2$, $R_\text{H}$ is very sensitive to the pressure [Fig. 4(c)] and the carrier doping [Fig. 4(d)]. The calculated $R_\text{H}$ is negative at low pressure (15 GPa, black line) and becomes positive at high pressures [30 GPa (red line) and 50 GPa (blue line)]. Moreover, a small amount of electron doping, such as 0.08 electron per formula, is able to induce a sign inversion of $R_\text{H}$. We also examined the magnetic field direction, which has minor influence on the sign of Hall resistance [Fig. 4(d)]. Such a behavior of the susceptible Hall resistance in the Pa-3 phase of NiTe$_2$ can be understood from its peculiar electronic structures. In comparison with the P-3m1 phase, there are less bands crossing the Fermi level [Figs. 3(c) and 3(d)] and just tiny Fermi pockets dispersing in the BZ [Figs. 3(e) and 3(f)] for the Pa-3 phase. This indicates that the intrinsic carrier concentrations of the Pa-3 phase are very low, which can be easily affected by external pressure or charge doping. The above results suggest that the experimentally observed sign reversal of Hall resistivity likely originates from the Pa-3 phase of NiTe$_2$.
%Furthermore, in our calculations the electronic relax time $\tau_{{\bf k}n}$ was considered as a constant and neglect its anisotropy of valence and conduction bands at different ${\bf k}$ points, which can also bring a obvious influence to the calculated $\rho_{xy}\tau$.
%The fact that the calculated sign of Hall resistance is inert in the P-3m1 phase but sensitive to many factors in the Pa-3 phase suggests that the observed sign inversion of Hall resistivity is likely originated from the Pa-3 phase of NiTe$_2$.

\section{Summary}

To summarize, to clarify the origins of the observed pressure-induced superconductivity\cite{presssc} and the sign reversal of Hall resistivity\cite{presshr} in NiTe$_2$, we have theoretically studied the crystal structure, electronic structure, electron-phonon coupling, and transport properties of NiTe$_2$ under several pressures. Our calculations show that the pressure can transform NiTe$_2$ from a layered P-3m1 phase to a cubic Pa-3 phase at $\sim$10 GPa. By comparing with the measured XRD spectra, we suggest that the realistic NiTe$_2$ compound may be a mixture of the P-3m1 and Pa-3 phases at high pressure. Our calculations also indicate that NiTe$_2$ retains the nontrivial topological property across the structural transition. The calculated superconducting $T_c$'s of the P-3m1 and Pa-3 phases both approach 0 K at 50 GPa, suggesting that the previously observed superconductivity in polycrystalline NiTe$_2$ may not be an intrinsic property. Due to the particular electronic structure, the calculated sign of the Hall resistance of the Pa-3 phase, in contrast to that of the P-3m1 phase, is very sensitive to the pressure and charge doping, which may be the origin of the observed sign reversal of Hall resistivity\cite{presshr}. Our calculation results on NiTe$_2$ wait for future experimental verification.

\begin{acknowledgments}

This work was supported by the National Key R$\&$D Program of China (Grants No. 2017YFA0302903 and No. 2019YFA0308603), the National Natural Science Foundation of China (Grants No. 11774424, No. 11774422, and No. 11734004), the Beijing Natural Science Foundation (Grant No. Z200005), the CAS Interdisciplinary Innovation Team, the Fundamental Research Funds for the Central Universities, and the Research Funds of Renmin University of China (Grant No. 19XNLG13). Computational resources were provided by the Physical Laboratory of High Performance Computing at Renmin University of China.

\end{acknowledgments}

\end{document}